# Noise-based communication and computing

Laszlo B. Kish

*Texas A&M University, Department of Electrical and Computer Engineering, College Station, TX 77843-3128, USA*

(First draft, August 18, 2008)

We discuss the speed-error-heat triangle and related problems with rapidly increasing energy dissipation and error rate during miniaturization. These and the independently growing need of unconditional data security have provoked non-conventional approaches in the physics of informatics. Noise-based informatics is a potentially promising possibility which is the way how biological brains process the information. Recently, it has been shown that thermal noise and its electronically enhanced versions (Johnson-like noises) can be utilized as information carrier with peculiar properties. Relevant examples are *Zero power (stealth) communication*, *Unconditionally secure communication with Johnson(-like) noise and Kirchhoff loop* and *Noise-driven computing*. The zero power communication utilizes the equilibrium background noise in the channel to transfer information. The unconditionally secure communication is based on the properties of Johnson(-like) noise and those of a simple Kirchhoff's loop. The scheme utilizes on the robustness of classical information and the second law of thermodynamics. It uncovers active eavesdropping within a single clock period (no error statistics is required) and it is naturally protected against the man-in-the-middle attack. Even though a practical system is never ideal, the engineering is straightforward to make the practical raw-bit security greater than quantum raw-bit security. The first practical realization of the communicator device have recently been reported and it consists of two computer cards and it is successfully tested beyond the quantum range, with parameters up to the range of 2000 km. Further advantages of the scheme is that the circuitry can easily be integrated on computer chips, unconditionally secure computer processors, memories and other hardware can be realized. We will also address noise driven logic offers lower energy dissipation for computing and shows interesting new properties to be briefly discussed during the talk.

## 1. Introduction

Current problems with rapidly energy dissipation and error rate during miniaturization and the growing need of unconditional security have provoked non-conventional approaches in the physics of informatics. Interestingly "noise" (stochasticity) is a common issue at all these aspects including the problem areas and the nonconventional ways of fighting these problems. Ferdinand Peper [1] has classified these efforts about the attempts of handling the noise issue:

*i.* Suppress it.

*ii.* Correct the errors caused by it.

*iii.* Live with it (in a clever way).

*iv.* Exploit it (such as biology does it).

Suppressing the noise (*i*) has only almost exhausted possibilities. The miniaturization of microelectronics implies and increasing noise as the result of the speed-error-heat triangle which indicates that we much give up at least one of these factors if we want to keep the other factor(s) in a good shape during miniaturization [2]. Smaller device means higher noise bandwidth and noise amplitude; both are resulting in greater error probability [2], see Figure 1.

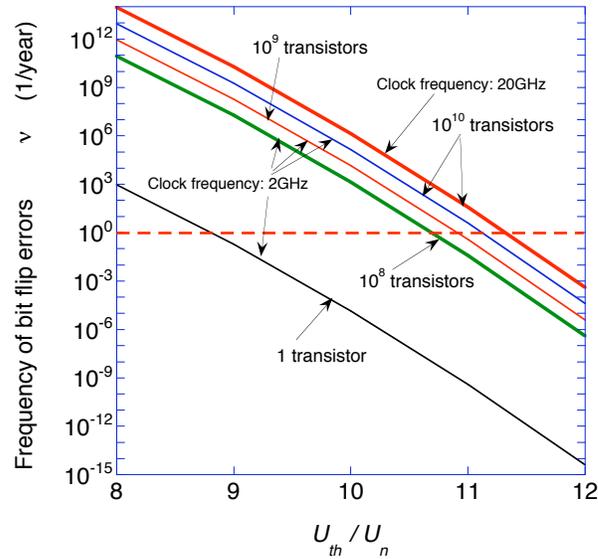

**Figure 1.** Error rate versus the ratio of noise margin (logic threshold) and RMS noise amplitude evaluated with the Rice formula of threshold crossing [2]. The error rate shows exponential growth with shrinking noise margin/noise ratio.

Moreover, smaller device size means lower supply voltage (to keep dissipation and electrical field limited) and that results in a narrower noise margin. All these effects cumulatively limit Moore's law at the statistical physical side [2], see Figure 2.

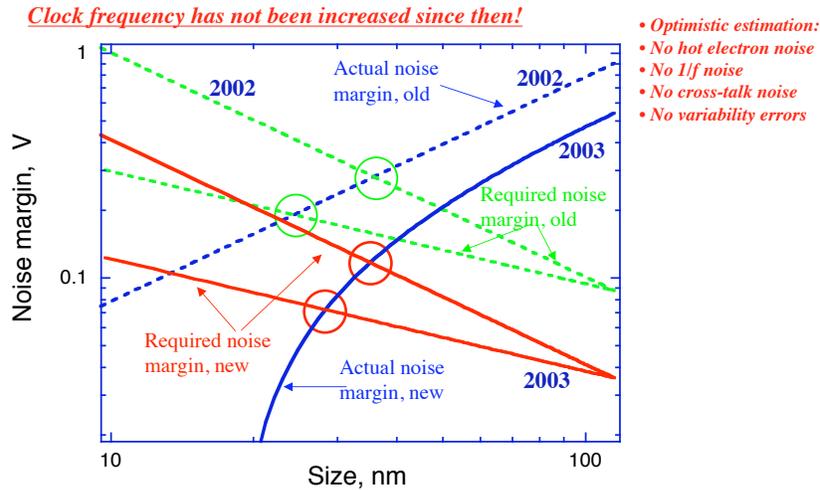

**Figure 2.** Noise margin due to dissipation/electrical field constraint and required noise margin to stay below 1 error/year due to thermal noise. Original estimation (2002, dashed lines [2]) and refined estimation (2003) [3]. The real evolution of required noise margin is between the estimation limits shown [2].

The predictions shown in Figure 2 neglected the leakage current and similar problems [3] and they supposed that the clock frequency in computers would have been kept growing as much as the increasing bandwidth with shrinking device size (and capacitances) allows. However, since the publication of [2], interestingly, the further *increase of the clock frequency was stopped* (usually stays around 3 GHz or much below). This situation does not change the required noise margin results shown in Figure 2, however it improves the situation at the dissipation constraint side but not at the electrical field constraint side. The result is managing dissipation better during miniaturization however serious problems with reliability and lifetime can occur due to increased electrical field. Thus Moore's law might be followed below the $\approx 30$ *nm* limit indicated by Figure 2 however a heavy toll can be expected on reliability/lifetime which however may turn out to be very profitable for the microprocessor industry (imagine processor replacement needs in the 1-2 year time scale like with light bulbs) even though very annoying for the user.

Fundamental studies [4,5] indicate that even quantum computers cannot help, whenever they are used as general-purpose machines, because their *Joule/bit* energy dissipation limit is about 1000 times greater [3] than that of classical computing. Neither processors built of single electron transistors can be expected to provide better results [6] unless the quantum dot size goes to the sub-nanometer range which would be an extraordinarily heavy toll on reliability and lifetime.

*Error correction* (*ii*), which is an old field or research and has a huge literature [1], is increasing the energy dissipation per bit and, in the case of large error probability, miniaturization cannot be based on it because it is more efficient to stay with higher characteristic device size in the chip than to miniaturize and give up speed and energy constraints due to error correction.

*Living with the noise* (*iii*) *without error correction means* either cleverly giving up accuracy and/or using proper coding or other trick may provide satisfactory information processing with much less energy dissipation [7-16] than normally. Here we should make a warning: one of these directions is the field of reversible computing, see for example in [14-17], which is a deadlock, according to the thorough analysis of Porod and coworkers [18-20] and Cavin and coworkers [21], because logical reversibility does not imply physical reversibility. Indeed, even the most elementary physical operation is irreversible at the thermodynamic level.

*Exploiting the noise* (*iv*) means, see for example [16,22-24] means that we are trying to make use of the noise as an information carrier, either in a computer, or in communication. This may be not a trivial task but challenging and promising because the brain signals are noises, thus the brain is utilizing such a trick.

The rest of this paper will briefly discuss some of the author's own initiatives to exploit the noise as information carrier in computation and data communication.

## 2. Stealth (classical and quantum) communication by noise

Recently, it has been shown [25,26] that the equilibrium thermal noise in information channels can be utilized to carry information. In this case, the transmitter does not emit any signal energy into the channel however it only modulates the existing noise there.

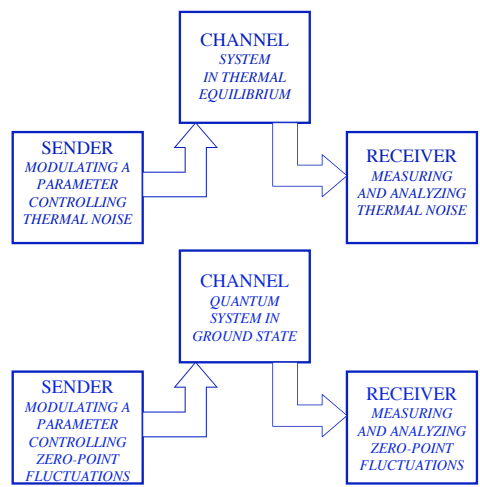

**Figure 3.** Stealth communications. Zero (signal) power classical communication (left) and zero-quantum quantum communication (right).

This issue is completely different/independent from the Porod-Landauer debate about the question if communication without net energy cost is possible by gaining back the energy spent in the communicator devices (even though Porod [20] is right). In our system, the noise is used as information carrier and no effort is made to restore the energy dissipated in the communicator devices. Therefore, this communicator is *not energy-free communication* but *it is free of emitted signal energy*. Zero (signal) power classical communication can utilize the modulation of background thermal noise in the information channel and zero-quantum quantum communication can utilize the modulation of the zero-point fluctuations in the quantum channel, see Figure 3, [25].

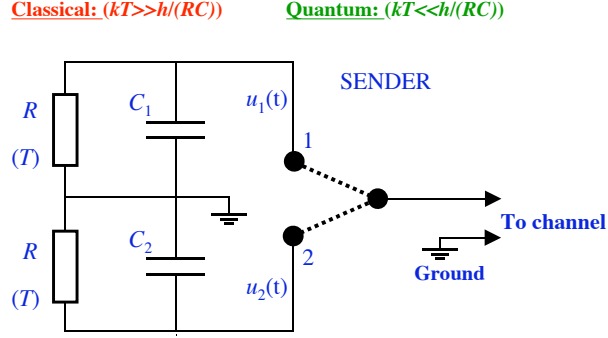

**Figure 4.** A possible realization of stealth communication with zero power classical communication or zero-quantum quantum communication utilizing classical thermal noise and zero-point quantum fluctuations with frequency bandwidth modulation. Classical or quantum, it depends on the upper cutoff frequency $f_c$ and the temperature. Classical: $hf_c \ll kT$; quantum: $hf_c \gg kT$. The receiver can be a simple noise spectrum analyzer or just a simple AC voltmeter.

In [25] some possible realizations were shown. Figure 4 shows one of the examples where the Johnson noise of resistors and bandwidth modulation are used in the classical and the quantum limits [25]. In the classical limit, $kT \gg h/RC$, the Johnson noise voltage spectrum is:

$$S_{u,class}(f) = 4kT \, \text{Re}[Z(f)] \, , \tag{2.1}$$

and in the quantum limit, $kT \ll h/RC$, the zero point voltage noise spectrum is:

$$S_{u,quan}(f) = 2hf \, \text{Re}[Z(f)] \, . \tag{2.2}$$

The contacted impedance $Z(f)$ is different in the two positions of the switch because of the different capacitance values. The receiver can be a simple noise spectrum analyzer or just a simple AC voltmeter.

In conclusion, it is possible to execute electronic data communication without injecting signal energy in the information channel. Because this the most invisible way of communication with basically background noise in the channel, it is proper to call these communications *stealth communication*, just like to call them zero-power classical communication or zero-quantum quantum communication.

## 3. Noise driven computing?

Because thermal noise turned out to be a special information carrier with low energy density in the information channel (even though the devices dissipate huge energy), see above [25,26], it is natural to pose the question: can we use it for information processing and computing? At thermal noise driven computing [24], an idea further inspired by the fact that the neural signals in the brain are noise, certain statistical parameters of the thermal noise is the information carrier.

The information may be carried by the bandwidth as at the zero power communication examples above or in other way. To study the minimal energy requirement of requirement/dissipation, another specific realization was used, a simple digital system with zero threshold voltage, when the thermal noise is equal to or greater than the digital signal [24]. Under these conditions, when the digital signal amplitude is less than the variance of the noise, classical digital information is usually considered to be zero or useless, see Figure 3.1.

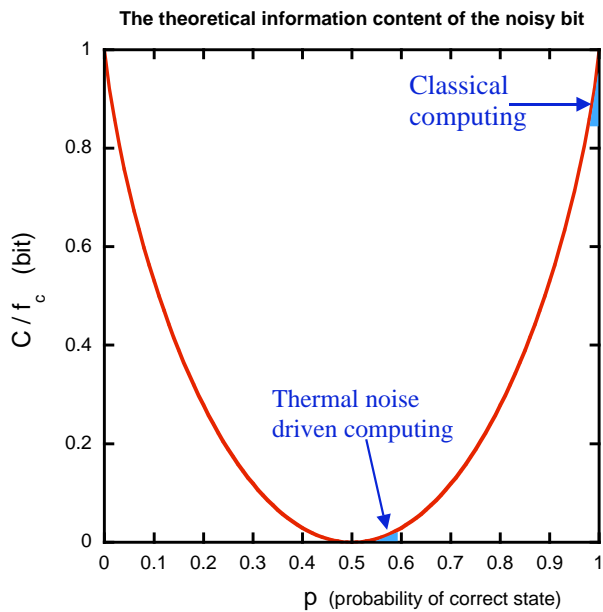

**Figure 5.** Illustration of the range of error probabilities of regular computing and that of the noise driven computing at a time moment. Time average or correlation analysis may show different result for the noise driven case. Y axies: bit content/single clock cycle; plot of Shannon's digital channel coding theorem. A noise driven computer will supposedly work in that range where the information content/ single switching is low, the signal looks like noise and no usual classical computer could work.

As we already mentioned in the Introduction related to point *iii* that, in a different class of efforts many notable scientists have been proposing efficient ways of working with probabilistic switches, which are noisy digital logic units with relatively high error probability. However, though these approaches can be relevant to future developments of thermal noise driven computing, they are very different from our present approach. In a *noise driven computer*, the voltage in the channel is fully or dominantly noise and the modulation of the statistical properties of this noise carry the information. Therefore, the way of extracting the information is not by "*error correcting the information*" like the efforts mentioned above do but by "*decoding the information in the noise*". The realization example we analyzed is working in the regime of huge error probability, $p \approx 0.5$, with zero logic threshold voltage and in the sub-noise signal amplitude limit; and all these parameters are very unusual.

Even though that at this stage there are more open questions than answers and we were are at the moment unable to show a functioning thermal noise driven logical circuitry, certain questions can already be answered about it. For the simplest case where the DC-shift-asymmetry of the amplitude distribution function of thermal carries the information, that is *small DC signal buried by thermal noise*, estimations were given about the information channel capacity and the minimal energy dissipation $E_1$ given in *Energy/bit* unit which is given as:

$$\eta = \frac{P_s}{C_{dig}\big|_{p \approx 0.5}} = \frac{\pi \ln 2}{2} kT / bit \approx 1.1 \ bit / kT \qquad (3.1)$$

where $P_s$ is the power needed to run the sub-noise level digital signal, $C_{dig}\big|_{p \approx 0.5}$ is he information channel capacity of this signal. This is a clear indication that noise has a potential as information carrier at computing.

The main advantage of such a hypothetical thermal noise driven computer would be a potentially improved energy efficiency and an obvious lack of leakage current, cross-talk and ground EMI problems due the very low DC voltages. An apparent disadvantage is the large number of extra (redundancy) elements required for error reduction. However, the above consideration was only to provide a lower limit of dissipation and the way noise can carry information can be very different from simply changing its mean value.

From the long list of open questions, we list some of the most important ones:

1. Do we need error correction at all, *except input/output operations*, in such a computer or when we want to simulate the way the brain works?

2. Is there any way to realize a non-Turing machine with stochastic elements without excessive hardware/software based redundance?

3. How should redundancy and error correction be efficiently used to run the thermal noise driven computer as a Turing machine?

4. How much energy would such error correction cost?

5. How much energy is needed to feed the active devices processing and producing the sub-thermal noise signals or the bandwidth control of thermal noise?

6. What is the impact of the internal noise of these devices?

Though, all these questions are relevant for the ultimate energy dissipation of thermal noise driven computers, the lower limit given for this specific arrangement with DC signal buried by thermal noise stays valid because this is the minimal energy needed to generate itself the digital signal.

**4. Unconditionally secure communication driven by noise [27]**

In today's software-based secure communications (tools we use when connect our bank via the internet), before the secure data exchange can start, the two communicators (Alice and Bob) must generate and share a joint secret (secure) encryption key through the communication channel while the eavesdropper (Eve) is supposedly monitoring the related data (Figure 6). This is a mathematically impossible task with current software methods thus they are only "computationally safe" that is Eve can decode the data but it takes too long time. Thus, if Eve had a genuine powerful algorithm or a sufficiently fast computer with standard algorithms, she could extract the secure key and decrypt the communicated data with a reasonable speed. Because new algorithms and computing solutions are continuously researched, today's software-based secure communication is a potential time bomb.

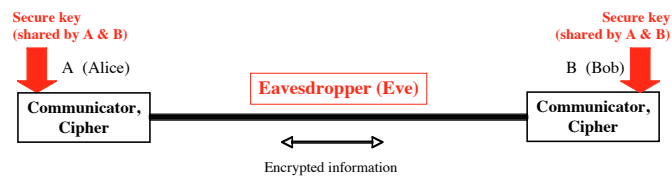

**Figure 6.** The two communicators (Alice and Bob) must generate and share a joint secure key through the communication channel while the eavesdropper (Eve) is monitoring the related data. This is an impossible task with software methods thus currently used software methods are only "computationally safe", which means they are potential time bombs.

Quantum key distribution (Stephen Wiesner 1970's; Charles H. Bennett and Gilles Brassard 1984; Artur Ekert 1990) has offered a solution which is claimed to be unconditionally secure [28-30]. The information bits are carried by single photons (Figure 7). Here the *no-cloning-theorem* of quantum physics is the theoretical foundation of security. It means that a single photon cannot be copied without noise (error). If Eve captures and measures the photon, it gets destroyed and she must regenerate and reinject it into the channel otherwise this bit will be considered invalid by Alice and Bob. However, due to the no-cloning rule, while Eve is doing that, she introduces noise and the error rate in the channel will become greater than without eavesdropping. Therefore, by evaluating the error statistics, after analyzing a number of transmitted bits and their errors, Alice and Bob will discover the eavesdropping. However, no quantum communicator is secure against the advanced type of the man-in-the-middle-attack, where Eve breaks the channel and installs two quantum communicators. With one of them she will communicate with Alice and pretends that she is Bob, and with the other one she will communicate with Bob and pretends that she is Alice. This is one example, where the secure wire communicator described below is superior to quantum encryption.

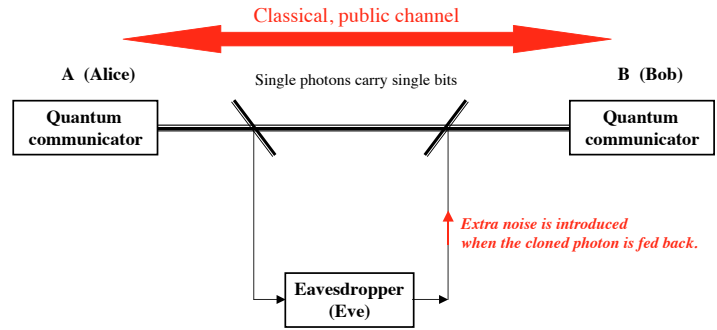

**Figure 7.** Generic quantum communication arrangement. To detect the eavesdropper, a statistics of bit errors must be built. That requires a sufficiently large number of bits. The communication of a few bits is not secure.

In 2005, an unconditionally secure classical-physical communication scheme, the Kirchhoff-loop-Johnson(-like)-Noise (KLJN) communicator [31] has been proposed, which is a statistical-physical competitor of quantum communicators [32,33]. It contains two identical pairs of resistors (Figure 3). The logic-low (L) and logic-high (H) resistors, $R_0$ and $R_1$, are randomly selected at the beginning of each clock period and are driven by their own Johnson-like noise (thermal noise) voltages or alternatively by their electrically enhanced versions with a pre-agreed factor (Johnson-like noise). The practical realizations [26, 34-36] contain more elements, such as filters, amplitude control units, etc. A secure key bit is generated and exchanged when the resistor values at the two ends differ. The role of the thermal noise is the measurement of the total resistance in the loop without serving information to Eve about the actual location of $R_0$ and $R_1$. When the total loop resistance gets known (this information is available to the public, too) and when it is the sum of $R_0$ and $R_1$, Alice and Bob can calculate the resistance value at the other side since they know their own resistance values.

The security of the communicator is based on two laws of physics (out of the rules of algebraic equations):

*a)* The *robustness of classical information* (just like quantum security is due to the fragility of quantum information) because this robustness allows the continuous monitoring of the signal;

b) The *second law of thermodynamics* (the *impossibility of constructing a perpetual motion machine*) [31] because that law guarantees that the net power flow between the resistors is zero in thermal equilibrium.

The KLJN scheme in Figure 8 is naturally protected [34] against the man-in-the-middle attack, and the active eavesdropping is detected immediately [31], within much shorter period than the time needed to transfer a single bit. Statistics of bit errors is not needed. The communication of even a single bit is secure [31]. Unconditionally secure networks can be easily realized and each station is using only two KLJN communicators to achieve that [35].

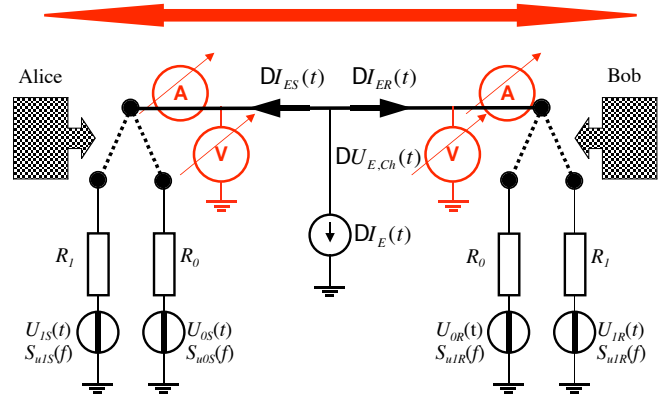

**Figure 8.** The KLJN wire communication arrangement. To detect the eavesdropper (represented for example by the current generator example at the middle), the instantaneous current and voltage data at measured at the two ends are broadcasted and compared. The eavesdropping is detected immediately, within much shorter period than the time needed to transfer a single bit. Statistics of bit errors is not needed. The communication of even a single bit is secure.

It is important to note that both the quantum and classical claims about unconditional security outlined above are valid only in idealized systems (at mathematical model level). In practical applications, no physical system is ideal and there are parasite effects and elements. Therefore, in practical applications, neither the quantum nor the KLJN schemes are totally secure. However, knowing their theory, their security and other performance can be designed depending on physical and financial limits. The ultimate test of security must be experimental: before marketing a secure communicator must be tested by all the known breaking methods.

Many quantum communicators have been reportedly built, up to the range of 200 km. Most of them are working through optical fibers and some of the most advanced and secure ones are able to communicate via air [30]. However, the experimental testing of different breaking ideas is more expensive than to build the communicators themselves! Therefore, because this technology is extremely expensive, the *quantum security against the various breaking attempts is mostly theoretical at the moment*, with a vast amount of theoretical-only papers about proposed breaking methods. It means that say about 20% of the necessary experimental work is done for the existing prototypes and, because there are many theoretical ways to break into a given quantum channel, the rest 80% of teh work must still be done before these quantum devices can be marketable at a large scale.

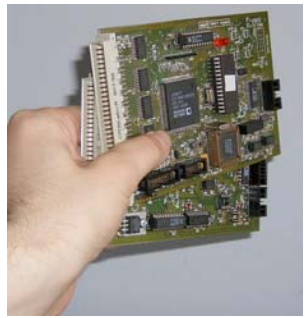

**Figure 9.** The KLJN wire communicator network element (communicator pair). Its fidelity is 99.98% and its security properties are superior to those of existing quantum communicators.

Due to the young age of the KLJN idea, so far, only one system has been built and tested [26,33,36] see Figure 9. It was experimentally verified against all the proposed attack types and, in all cases, the active eavesdropping was discovered during the communication of a single bit. The speed of this communicator is half of the speed of the most advanced quantum communicator [30] however there are straightforward ways to increase the speed in future designs. Its price was a few hundred dollars and, in an integrated form, its fabrication price will be similar to that of an Eternet card in a PC. The system is network ready and it is robust against vibrations, dust, ageing, etc. With proper filters, currently used wire likes, such as power lines, phone lines and internet lines can also be used as channels [37].

Finally, a potentially important application [38] is securing a laptop computer, a hardware or an algorithm. In the case of the need of extraordinary security, the KLJN communicator can easily be integrated on existing types of digital chips in order to provide secure data communication between hardware processors, memory chips, hard disks and other units within a computer or other data processor system. The secure key exchange can take place at the very first run and the system can renew the key later at random times with an authenticated fashion to prohibit man-in-the-middle attack. The key can be stored in flash memories within the communicating chip units at hidden random addresses among other random bits that are continuously generated by the secure line but are never actually used. Thus, even if the system is disassembled, and the eavesdropper can have direct access to the communication lines between the units, or even if she is trying to use a man-in-the- middle attack, no information can be extracted. The only way to break the code is to learn the chip structure, to understand the machine code program and to read out the information during running by accessing the proper internal ports of the working chips. However such an attack needs extraordinary resources and even that can be prohibited by a password lockout. The unconditional security of commercial algorithms against piracy can be provided in a similar way.